\documentclass{article}
\usepackage{amsmath}
\usepackage{amsfonts}
\usepackage{amssymb}
\usepackage{graphicx}%
\newcommand{\bpartial}{\mathop{\partial\kern -4pt\raisebox{.8pt}{$|$}}}
\newcommand{\bra}{\mathopen{[\kern-1.6pt[}}
\newcommand{\ket}{\mathclose{]\kern-1.5pt]}}
\newcommand{\bbra}{\mathopen{[\kern-2.2pt[\kern-2.3pt[}}
\newcommand{\bket}{\mathclose{]\kern-2.1pt]\kern-2.3pt]}}

\makeindex
\begin{document}

\title {\large{ \bf Exchanging role of the phase space and symmetry group of integrable Hamiltonian systems related to  Lie bialgebras of bi-symplectic types  }}
\vspace{3mm}
\author {\small{ \bf J. Abedi-Fardad$^{1}$ }\hspace{-2mm}{ \footnote{ e-mail: abedifardadjafar@yahoo.com}},
\small{ \bf A. Rezaei-Aghdam$^2$ }\hspace{-1mm}{\footnote{ e-mail: rezaei-a@azaruniv.ac.ir }},
\small{ \bf  Gh. Haghighatdoost$^1$}\hspace{-1mm}{ \footnote{ e-mail: gorbanali@azaruniv.ac.ir}}  \\
{\small{$^{1}$\em Department of Mathematics, Faculty of science, Azarbaijan Shahid Madani University, 53714-161, Tabriz, Iran}}\\
{\small {$^{2}$\em Department of Physics, Faculty of science, Azarbaijan Shahid Madani University, 53714-161, Tabriz, Iran}} }
  \maketitle

\begin{abstract}
 We construct integrable Hamiltonian systems with Lie bialgebras $({\bf g} , {\bf \tilde{g}})$ of the bi-symplectic type  for which the Poisson-Lie
 groups ${\bf G}$ play the role of the phase spaces, and their dual Lie groups ${\bf {\tilde {G}}}$ play the  role of
the symmetry groups of the systems. We give the new transformations to exchange the role of phase spaces and symmetry groups and obtain the relations between integrals of motions of these  integrable systems. Finally, we give some examples of real four-dimensional Lie bialgebras of bi-symplectic type.
\end{abstract}

{\bf keywords:}{ Integrable  Hamiltonian systems,  Lie bialgebra, Bi-symplectic structure.}

\section {\large {\bf Introduction}}

A Hamiltonian system with $N$ degrees of freedom is integrable in the sense of  Liouville theorem  if it has $N$
invariants  (globally defined and functionally independent)  in involution (for review see \cite{STSM}, \cite{AVB}, \cite{OBD}). If an integrable Hamiltonian system is
invariant under some transformations on the phase space variables (such that these transformations construct a group); then we have
 a symmetry group for this Hamiltonian system \cite{OBD}. In most of the symmetric integrable Hamiltonian systems, the symmetry
group is a Lie group. A Lie group with compatible Poisson structure on it; is called Poisson-Lie group \cite{Drin}.
Poisson-Lie groups and their algebraic forms i.e. Lie bialgebras play important roles in constructing of classical integrable
Hamiltonian systems \cite{YKS}. In \cite{JAF}, we have  constructed integrable
and superintegrable Hamiltonian systems for which the symmetry Lie groups are also the phase spaces of the systems.
Also in \cite{JAF2}, we  classified all four-dimensional real Lie bialgebras $({\bf g} , {\bf \tilde{g}})$  of symplectic type and also give two
examples as the physical applications, such that for
these integrable systems the Poisson-Lie groups ${\bf G}$ play the role of
the phase spaces, and their dual Lie groups ${\bf {\tilde {G}}}$  play the  role of
the symmetry Lie groups of the systems. In this work,  we give  new transformations for exchanging
 the role of the phase space and the symmetry Lie group. We use  Lie bialgebras of
  bi-symplectic types for constructing of the integrable Hamiltonian systems for which the Poisson-Lie groups
  ${\bf G}$ play the role of the phase spaces, and their dual Lie groups ${\bf {\tilde {G}}}$ play the  role of the
symmetry groups of the systems; then using these of transformation, the role of  ${\bf G}$  and ${\bf {\tilde {G}}}$ will be
exchanged.
The outline of the paper is as follows. For self-containing of the paper, we review the construction of an integrable Hamiltonian system
by using  Lie bialgebras in section 2. Then, in section 3, we consider integrable Hamiltonian systems related to Lie
bialgebras of bi-symplectic types and present  new transformations that exchange the role of phase spaces and symmetry Lie groups.
In section 4, we give some examples related to real four-dimensional Lie bialgebras of bi-symplectic types  \cite{JAF2}.

\smallskip

\section {\large {\bf   Review of the construction of integrable Hamiltonian system with Lie bialgebra}}

For introducing the notations, let us have a survey on  some definitions about Lie bialgebras \cite{Drin}
and related integrable Hamiltonian systems that can be constructed from them  (for a review see \cite{YKS}).

{\bf Definition}\cite{Drin}: A {\em Lie bialgebra} is a Lie algebra ${\bf g}$ with a skew-symmetric linear map
$\delta : \bf g \rightarrow  \bf g\otimes \bf g$ such that:\\
a) $\delta$ is a one-cocycle, i.e.:
\begin{equation}
\delta ([X,Y])=[\delta (X),1\otimes Y+Y\otimes 1]+[1\otimes X+X\otimes 1,\delta(Y)],\;\;\;\;\;\;\;\;\forall X,Y\in \bf g,
\end{equation}
where 1 is the identity map on $\bf g$.\\
 b) ${\delta}^t :\tilde{\bf g} \otimes \tilde{\bf g} \rightarrow \tilde{\bf g}$ is a Lie
bracket on $\tilde{\bf g}$ ( $\tilde{\bf g}$ is the dual space of the vector space $\bf g$ ):
\begin{equation}\label{BA6}
(\xi\otimes \eta ,\delta(X))=({\delta}^t(\xi\otimes \eta),X)=([\xi ,\eta]_{\tilde{\bf g}} ,X),\;\;\;\;\;\;\;\;\forall X\in \bf g;\;\; \xi,\eta\in \tilde{\bf g},
\end{equation}
where $( .,. )$ is a standard inner product between $\bf g$ and $\tilde{\bf g}$.
 The Lie bialgebra defined in this way will be denoted by $(\bf g,\tilde{\bf g})$ or $(\bf g ,\delta)$ .
In terms of $\{X_i\}$ and $ \{\tilde{ X}^j\}$ (the bases of the Lie algebras $\bf g$ and $\tilde{\bf g}$ respectively), we have
the following commutation relations \cite{YKS} :
\begin{eqnarray}\label{A41}
[X_i , X_j] = f_{ij}^{\;\;\;k} X_k, \hspace*{1cm} [\tilde{X}^i ,\tilde{ X}^j] ={{\tilde{f}}^{ij}}_{\; \; \: k} {\tilde{X}^k},\nonumber\\
\lbrack X_i,\tilde{X}^j\rbrack ={\tilde{f}^{jk}}_{\; \; \; \:i} X_k + f_{ki}^{\;\;\;j} \tilde{X}^k,\hspace*{1.5cm}
\end{eqnarray}
such that $\delta(X_i) = {\tilde{f}^{jk}}_{\; \; \; \:i} X_j \otimes X_k $ and on the vector space
$\cal{D} = {\bf g}\oplus{\bf \tilde{g}}$ we have a Lie algebra structure and ad-invariant isotropic bilinear form $<.~ , ~. >$:
\begin{equation}\label{BB5}
<X_i , X_j> = <\tilde{X}^i , \tilde{X}^j> = 0, \hspace{10mm}
\;<X_i , \tilde{X}^j>\; ={\delta^j}_i.
\end{equation}
the  triple $(\cal{D} , {\bf g} , {\bf \tilde{g}})$ is called {\em Manin triple} \cite{YKS}.

For the coboundary Lie bialgebra, there is an  $r$-matrix ($r\in \bf g\otimes\bf g$) such that:
\begin{equation}\label{BA1}
\delta (X)=[1\otimes X+X\otimes 1,r ],
\end{equation}
if the $r$-matrix  $r=r^{ij}X_i\otimes X_j$ satisfy the classical Yang-Baxter equation (CYBE) \cite{MAS}:
\begin{equation}\label{4bb77}
[[r,r]]=[r_{12},r_{13}]+[r_{12},r_{23}]+[r_{13},r_{23}]=0 ;
\end{equation}
(where $r_{12}=r^{ij}X_i\otimes X_j\otimes 1$, $r_{13}=r^{ij}X_i\otimes 1
\otimes X_j$ and  $r_{23}=r^{ij} 1\otimes X_i\otimes X_j$); then the Lie bialgebra  $({\bf g} , {\bf \tilde{g}})$
is called triangular Lie bialgebra \cite{YKS}. The CYBE (\ref{4bb77}) can be rewritten in the term of
structure constants $f_{ij}^{\;\;\;k}$ as:
\begin{equation}\label{4bb1}
r^{ij} r^{kl} f_{ik}^{\;\;\;m}+r^{mi} r^{kl} f_{ik}^{\;\;\;j}+r^{mi} r^{jk} f_{ik}^{\;\;\;l}=0.
\end{equation}
Now, let us consider the method of constructing classical dynamical system which have symmetry Lie group.
Let us consider $2n$ dimensional manifold ${\bf M}$  with the symplectic structure $\omega _{ij}$ and local coordinate
 $\lbrace x_i \rbrace$
 as a phase space. The Poisson bracket $\lbrace .,.\rbrace$ (defined by symplectic structure $\omega _{ij}$) of arbitrary
functions $f,g \in C^{\infty}({\bf M})$  is given by
\begin{equation}
\lbrace f,g \rbrace ={\bf P}^{ij} \frac{\partial f}{\partial x_i}\frac{\partial f}{\partial x_j},
\end{equation}
where ${\bf P}^{ij}$ is the inverse of the matrix $\omega _{ij}$. For a dynamical system with symmetry Lie group
${\bf {G}}$
we have independent dynamical functions $S_k=S_k(x_i) ,  (k=1,...,dim({\bf g}))$ which are constructed
 as functions on ${\bf M}$  and satisfy the  following relations
\begin{equation}\label{4bb88}
\lbrace S_i,S_j \rbrace ={ f}_{ij}^{\;\;\;k} S_k,
\end{equation}
where ${ f}_{ij}^{\;\;\;k}$ are structure constants of the Lie algebra ${\bf {g}}$  of the symmetry Lie group
 ${\bf {G}}$ .
For specifying the dynamical system with symmetry group, there are two methods:

a) Consider an   $r$-matrix  related
to the Lie bialgebra $({\bf g} , {\bf \tilde{g}})$. If one can choose a matrix
 representation for the Lie algebra  ${\bf {g}}$;
then the satisfaction of the  ${\bf g}$-valued functions:
\begin{equation}\label{4bb3}
Q=S_i { r}^{ij}{ X}_j,
\end{equation}
in the relation:
\begin{equation}
\lbrace Q\stackrel{\otimes}{,}Q \rbrace + [ Q\otimes I+I \otimes Q, r] =0,
\end{equation}
is equivalent to the CYBE (\ref{4bb1}) for the $ r$-matrix \cite{RBZ}.
Then, one can see that the following functions are the constants of motion of a  dynamical system \cite{RBZ}:
\begin{equation}\label{4bb331}
I_k=trace(Q^k), \hspace*{1cm} k\in \mathbb{N}.
\end{equation}

b) For the case that there is no relevant matrix representation of the Lie algebra ${\bf g}$; one can use a less accurate method;
so that after writing relations (\ref{4bb88}); one can specify the maximal number of the dynamical functions $S_i$ which are
in involution i.e. $\lbrace S_i,S_j \rbrace =0$;  then one can consider one of those functions $S_i$ as a Hamiltonian
of the system \cite{OBD}.

We know that a dynamical system with $n$ degrees of freedom would be completely integrable (in the sense of Liouville theorem)
if there are $n$ independent
 constants of motion. In the next section, we will construct integrable systems by using Lie  bialgebra $({\bf g} , {\bf \tilde{g}})$ of symplectic type  such that for these systems
 the phase spaces are symplectic Poisson-Lie
  groups ${\bf G}$  for which the related dual Lie groups ${\bf \tilde{G}}$ are symmetry Lie groups and vice versa.
In section four we will consider examples related to real four-dimensional Lie bialgebras of bi-symplectic type \cite{JAF2}.

\section {\large {\bf   Integrable Hamiltonian systems related to Lie bialgebras of bi-symplectic type, exchanging the role of
phase spaces and symmetry Lie groups }}

Let us first consider the Lie bialgebra of bi-symplectic type and as a first step we consider the definition of the symplectic structure on
a Lie algebra \cite{GO}.

{\bf Definition}\cite{GO}: A two form $\omega \in ({\bf \tilde{g}} \wedge {\bf \tilde{g}} ) $ on a Lie algebra ${\bf g}$ is
called symplectic if $d\omega \in \wedge ^3 {\bf \tilde{g}}$
\begin{equation}
d\omega (X,Y,Z)=-\omega ([X,Y],Z)+\omega ([X,Z],Y)-\omega ([Y,Z],X), \hspace*{2cm} \forall X,Y,Z\in {\bf g},
\end{equation}
is closed i.e. $d\omega =0$ and $\omega $ is nondegenerate{\footnote{Note that here ${\bf \tilde{g}}$ is the dual
 space of ${\bf g}$ and $(.,.)$ is the standard inner product between ${\bf {g}}$  and ${\bf \tilde{g}}$.  }} . In terms of Lie algebra basis
 $\omega = \omega _{ij} \tilde{X}^i\wedge \tilde{X}^j  $ we must have:
\begin{equation}
f_{ij}^{\;\;\;l} \omega _{lk}+f_{ik}^{\;\;\;l} \omega _{lj}+f_{jk}^{\;\;\;l} \omega _{li}=0,
\end{equation}
the inverse of $\omega _{ij}$ can be considered by $P^{ij}$  as Poisson structure on a Lie algebra ${\bf g}$. The Poisson
structure on the corresponding Lie groups ${\bf G}$ can be obtained by using of vielbeins $e^i_{\;j}$ \cite{MNA}
\begin{equation}
{\bf P}^{ij}(x)= e^i_{\;k} e^j_{\;l} P^{kl},
\end{equation}
such that
\begin{equation}
dg g^{-1}= e_i^{\;j} dx^i X_j , \hspace*{2cm}\forall g\in {\bf G},
\end{equation}
and
\begin{equation}
 e^i_{\;j} e_k^{\;j} =\delta ^i _k,
\end{equation}
where $ e_k^{\;j} $ is inverse of $e^i_{\;j}$
and $\{x^i\}$ and $\{X_i\} $ are coordinates and generators of the Lie group ${\bf G}$.
In the same way, if one of the Lie algebras ${\bf g}$ or ${\bf \tilde{g}} $ from Lie bialgebra
$({\bf g},{\bf \tilde{g}})$ is of the symplectic type, then it is called symplectic Lie bialgebra. In the case that both of
${\bf g}$ and ${\bf \tilde{g}} $ of $({\bf g},{\bf \tilde{g}})$ are of the symplectic type, then the Lie bialgebra $({\bf g},{\bf \tilde{g}})$
is called  bi-symplectic type \cite{JAF2}. Note that for these Lie bialgebras of bi-symplectic type  the dimension of Lie algebra
${\bf g}$ and ${\bf \tilde{g}} $ must be even $(2n)$.
 In \cite{GO} all real four-dimensional  Lie algebras of symplectic type are
classified; also all real four-dimensional Lie bialgebras of symplectic and bi-symplectic types are classified in \cite{JAF2}.

Now, we construct a dynamical system by using of  a $2n$ dimensional real  Lie bialgebra $({\bf g},{\bf {\tilde{g}}})$ of
 bi-symplectic type (for which  Lie algebras  ${\bf g}$ and ${\bf {\tilde{g}}}$  are isomorphic) such that
for this dynamical system  the Poisson-Lie group
  ${\bf G}$ (with coordinates ($x^1$, ..., $x^{2n}$))  plays the role of the phase space, and its dual Lie group ${\bf {\tilde {G}}}$
   (with coordinates ($y^1$, ..., $ y^{2n}$) )  plays the  role of symmetry Lie group, so we have
\begin{equation}\label{1A51}
\lbrace S^i(x^1,...,x^{2n}),S^j(x^1,...,x^{2n}) \rbrace =\tilde{f}^{ij}_{\;\;\;k} S^k(x^1,...,x^{2n}) .
\end{equation}
Now,  by using   the assumption that ${\bf g}$ and ${\bf {\tilde{g}}}$ are isomorphic i.e., there is a matrix
$C$ for which:
\begin{equation}\label{4bb2}
\tilde{X}^i=C^{il} X_l  \;,\;\;[\tilde{X}^i,\tilde{X}^j]=\tilde{f}^{ij}_{\;\; \;  k} \tilde{X}^k,
\end{equation}
  then we have
\begin{eqnarray}\label{5A1}
{\tilde{f}}^{ij}_{\;\; \;  k}&=&C^{il} C^{jm} f_{lm}^{\;\;\;s} (C^{-1})_{sk},
\end{eqnarray}
where $C^{ij}$ is  an invertible isomorphism matrix; such that by
 substituting   (\ref{5A1}) in (\ref{1A51}) and defining
 \begin{eqnarray}\label{5A2AA}
 {\tilde S} _m=(C^{-1})_{ml}S^l ,
 \end{eqnarray}
  we have
\begin{eqnarray}\label{5A2}
\lbrace {\tilde S} _l(x^1,...,x^{2n}),{\tilde S} _m(x^1,...,x^{2n})\rbrace = f_{lm}^{\;\;\;k} {\tilde S}_ k(x^1,...,x^{2n}).
\end{eqnarray}

On the other hand, one can consider the coordinates $\{x^i\}$ on a Poisson-Lie group ${\bf G}$ (as a  phase space with Poisson structure ${\bf P}^{ij}$)
as functions of the coordinates $\{y^i\}$ of the Poisson-Lie group ${\bf {\tilde {G}}}$.{\footnote{Note that this transformation
is not (in general) canonical transformation.} }So, on a symplectic Poisson-Lie group ${\bf {\tilde {G}}}$ we have:
\begin{eqnarray}\label{A3}
\lbrace x^i(y^1,...,y^{2n}),x^j(y^1,...,y^{2n})\rbrace= {\bf \tilde{P}}^{lk}\;.\frac{\partial x^i(y^1,...,y^{2n})}{\partial y^l}\;.\frac{\partial x^j(y^1,...,y^{2n})}{\partial y^k}={\bf P}^{ij},
\end{eqnarray}
where the functions ${\tilde S} _i(x^1(y^1,...,y^{2n}),...,x^{2n}(y^1,...,y^{2n}))={\tilde S} _i(y^1,...,y^{2n})$ can be considered as  dynamical
functions on the space  ${\bf {\tilde {G}}}$ i.e. relation (\ref{5A2}) tells us that under transformations (\ref{A3})
$x^i=x^i(y^1,...y^{2n})$ the role of phase space ${\bf G}$ and symmetry Lie group  ${\bf {\tilde {G}}}$  is exchanged.
Furthermore, using (\ref{4bb3}), (\ref{4bb2}) and definition of  ${\tilde S}_i$ (\ref{5A2AA}), one can show that the
${\bf {\tilde{g}}}$-valued functions $Q$ (\ref{4bb3}) is transformed to ${\bf {{g}}}$-valued functions ${\tilde{Q}}$ as follows
\begin{eqnarray}\label{4bb99}
Q(x^1(y^1,...,y^{2n}),...,x^{2n}(y^1,...,y^{2n}))&=&S^i(x^1(y^1,...,y^{2n}),...,x^{2n}(y^1,...,y^{2n})) {\tilde{r}}_{ij}{\tilde{X}}^j \nonumber\\
&=&\tilde{Q}(y^1,...,y^{2n})=\tilde{S}_i(y^1,...,y^{2n}) {r}^{ij}{X_j},
\end{eqnarray}
if
\begin{eqnarray}\label{4bb1b}
\tilde{r}_{ij}=(C^{-1})_{ki} r^{kl} (C^{-1})_{lj},
\end{eqnarray}
where $\tilde{r}_{ij}$ and ${r}^{ij}$ are the solutions of CYBE for ${\bf {\tilde{g}}}$ and
${\bf g}$, respectively. Furthermore, one can show that (using (\ref{5A1}) and (\ref{4bb1b}) )  CYBE for
the Lie algebra ${\bf g}$ (\ref{4bb1}) transform the  CYBE  for the Lie algebra ${\bf {\tilde{g}}}$ and vice versa.
 In this way, we have the following theorem for exchanging  the role of the phase space   ${\bf G}$ and symmetry Lie group
 ${\bf {\tilde {G}}}$ and vice versa.\\

 {\bf Theorem 1:}
\textit{ Let $({\bf g},{\bf {\tilde{g}}})$ is a bi-symplectic Lie bialgebra for which the Lie algebras ${\bf g}$ and ${\bf {\tilde{g}}}$
 are isomorphic. The  dynamical functions  of the dynamical system  with Lie group
  ${\bf G}$ as a phase space and its dual Lie group ${\bf {\tilde {G}}}$ as symmetry Lie group are related to
   the dynamical functions  of the dynamical system
 with Lie group ${\bf {\tilde {G}}}$ as a phase space and  Lie group ${\bf G}$ as symmetry Lie group, as follows:
 \begin{eqnarray}\label{A5e1}
{\tilde S} _j(y^1,...,y^{2n}) =(C^{-1})_{jl}S^l(x^1(y^1,...,y^{2n}),...,x^{2n}(y^1,...,y^{2n})),
\end{eqnarray}
for which the transformation  $x^i=x^i(y^1,...,y^{2n})$ is a solution  of (\ref{A3}) and the matrix  $C^{ij}$ is the isomorphism matrix. Furthermore, the integrals of motion of these systems are related to each other by using  of  (\ref{A5e1}).}\\

In this way, we find  transformations (\ref{A3}) which change the role of phase space and symmetry Lie group.
 In the next section,  we will consider Hamiltonian dynamical systems related to real four-dimensional Lie bialgebras of bi-symplectic type  as
some examples.

 \section {\large {\bf Some examples }}
Now, we consider some integrable systems obtained by using the real four-dimensional
 Lie bialgebras of bi-symplectic type  \cite{JAF2}. In these examples, we consider the Lie group ${\bf G}$
related to the Lie bialgebra $({\bf g}, {\bf {\tilde{g}}})$ as a phase space, and its dual Lie group ${\bf {\tilde {G}}}$ as a
symmetry group of the system; such that  by using theorem 1 we exchange the role of phase space and symmetry group
and obtain other integrable system for which the Lie group ${\bf G}$ plays the role of symmetry Lie group  and its dual Lie group
${\bf {\tilde {G}}}$ plays the role of a phase space of the system, for this propose we use the formalisms
mentioned in the previous section  for calculation of   integrable Hamiltonian systems with some symmetry groups ${\bf {\tilde {G}}}$.
In example one, we use the method   a) and in the other examples 2-5 we use the  b) method.\\
 \smallskip

{\bf Example 1)} Lie bialgebra $(A_{4,9}^0.iv,A_{4,9}^0)$ \cite{JAF2}:\\
 Consider the Lie group $\bf{A_{4,9}^0.iv} $ as a phase space and ${\bf A_{4,9}^0}$ as symmetry Lie group of a Hamiltonian system.  For this example, the Darboux coordinates have the following forms \cite{JAF}:
 \begin{eqnarray}\label{A4555}
 z_1&=&e^{-x_1},\hspace*{3.5cm} z_2=\frac{e^{x_1}x_2}{e^{x_1}-1},\nonumber\\
  z_3&=&-\frac{e^{2x_1}(x_2+2x_4)}{2(-1+e^{x_1})},\hspace*{0.9cm}\;\;\;\;\;\;\;\; z_4=x_3 -\frac{ e^{-x_1}}{2},
 \end{eqnarray}
  such that in this Darboux coordinates the symplectic structure on the Lie group  $\bf{A_{4,9}^0.iv} $ which has the following form \cite{JAF2}:
 \begin{eqnarray}
 \lbrace x_1,x_4\rbrace =1-e^{-x_1},\;\;\lbrace x_2,x_3\rbrace =1-e^{-x_1},\;\;\lbrace x_2,x_4\rbrace =x_2e^{-x_1},\;\;\lbrace x_3,x_4\rbrace =\frac{1+e^{-2x_1}-2e^{-x_1}}{2},
 \end{eqnarray}
 can be simplified as
{\begin{equation}\label{4b2}
 \lbrace z_1,z_3\rbrace =1~,~~~~ \lbrace z_2,z_4\rbrace =1 .
\end{equation}
 In this way, we have the following forms for the dynamical functions $S^i$ according to \cite{JAF}
\begin{eqnarray}\label{2A4112}
S^1&=&- z_3=\frac{e^{2x_1}(x_2+2x_4)}{2(-1+e^{x_1})}, \nonumber\\
S^2&=&- z_4=-x_3 +\frac{ e^{-x_1}}{2},\nonumber\\
S^3&=&- z_2z_3= \frac{x_2e^{3x_1}(x_2+2 x_4)}{2(-1+e^{x_1})^2} ,\nonumber\\
S^4&=&-z_1 z_3-z_2 z_4=\frac{x_2(1+e^{x_1}(1-2x_3))+2e^{x_1}x_4}{2(-1+e^{x_1})} ,
\end{eqnarray}
 such that these functions  satisfy  the following Poisson brackets
\begin{eqnarray}
 \lbrace S^1,S^4\rbrace =S^1\;,\; \lbrace S^2,S^3\rbrace =S^1\;,\; \lbrace S^2,S^4\rbrace =S^2\;,
\end{eqnarray}
i.e., a Poisson bracket $ \lbrace S^i,S^j \rbrace ={\tilde f}^{ij}_{\;\;k} S^k  $, where ${\tilde f}^{ij}_{\;\;k}$ are the structure constants  of Lie algebra ${A_{4,9}^0} $.
We can take a representation of four-dimensional triangular matrices for the basis of the Lie algebra ${A_{4,9}^0} $ as follows:
\begin{equation}
{\tilde X}^1=\left(
\begin{matrix}
0&a&0&0\cr
0&0&0&0\cr
0&0&0&0\cr
0&0&0&0
 \end{matrix}
\right) ,\; {\tilde X}^2=\left(
\begin{matrix}
0&0&1&0\cr
0&0&0&0\cr
0&0&0&0\cr
0&0&0&0
 \end{matrix}
\right) ,\; {\tilde X}^3=\left(
\begin{matrix}
0&a b&0&c\cr
0&0&0&0\cr
0&a&0&0\cr
0&0&0&d
 \end{matrix}
\right) ,\; {\tilde X}^4=\left(
\begin{matrix}
0&0&b&0\cr
0&1&0&0\cr
0&0&1&0\cr
0&0&0&0
 \end{matrix}
\right),
\end{equation}
where $a, b, c$ and $ d $ are nonzero arbitrary real constants.
Then from (\ref{4bb3}) and (\ref{4bb331}) for the $r-$matrix \\
 $\tilde{r}=-1/2{\tilde X}^1\wedge {\tilde X}^2-{\tilde X}^1\wedge {\tilde X}^4-{\tilde X}^2\wedge {\tilde X}^3$, \cite{JAF2}
 the constants of motion are obtained as follows:
\begin{eqnarray}\label{4bb2}
 I_1=d z_4+2 z_3=d (x_3-\frac{e^{-x_1}}{2})+\frac{ e^{2x_1}(x_2+2x_4)}{1-e^{x_1}},\nonumber\\
 I_2=(d z_4)^2+2 (z_3)^2=(d (x_3-\frac{e^{-x_1}}{2}))^2+\frac{ e^{4x_1}(x_2+2x_4)^2}{2(e^{x_1}-1)^2}.
\end{eqnarray}
Note that in terms of the Darboux coordinates $z_i$ the constants of motion $I_1$ and $I_2$ can be related to the physical systems.
Now, by using (\ref{A3}) and Poisson structures on the Lie groups ${\bf A_{4,9}^0}$ and ${\bf A_{4,9}^0.iv}$   \cite{JAF2} one can find $x_i$'s as functions of $y_i$ as follows:
\begin{equation}\label{2A511}
x_1=y_4\;,\;\;x_2= y_3\;,\;\;x_3=-y_2+y_4\;,\;\;x_4=-y_1,
\end{equation}
For the Lie group ${\bf A_{4,9}^0}$ as phase space the  Darboux coordinates have the following forms \cite{JAF}:
 \begin{eqnarray}\label{2A511p}
{\tilde z_1}&=&\frac{e^{-y_4}}{2}+y_2-y_4,\hspace*{2.6cm} {\tilde z_2}=e^{y_4}+\frac{e^{2y_4}(-2y_1+y_3)}{2(e^{y_4}-1)},\nonumber\\
 {\tilde z_3}&=&\frac{e^{y_4}y_3}{e^{y_4}-1},\hspace*{3cm}\;\;\;\;\;\;\;\;{\tilde z_4}= e^{-y_4},
 \end{eqnarray}
  such that in this Darboux coordinates, the symplectic structure on the Lie group $\bf{A_{4,9}^0} $ which has the following form \cite{JAF2}:
 \begin{eqnarray}
 \lbrace y_1,y_2\rbrace =1/2(1-e^{-2y_4}),\;\;\lbrace y_1,y_3\rbrace =y_3e^{-y_4},\;\;\lbrace y_1,y_4\rbrace =1-e^{-y_4},\;\;\lbrace y_2,y_3\rbrace =1-e^{-y_4},
 \end{eqnarray}
 can be simplified as
{\begin{equation}\label{bb24}
 \lbrace {\tilde z_1},{\tilde z_3}\rbrace =1~,~~~~ \lbrace {\tilde z_2},{\tilde z_4}\rbrace =1 .
\end{equation}
Then by substituting (\ref{2A511}) in (\ref{2A4112})  and using (\ref{5A2AA}) with
\begin{equation}
C=\left(
\begin{matrix}
0&0&0&1\cr
0&0&1&0\cr
0&-1&0&1\cr
-1&0&0&0
 \end{matrix}
\right),
\end{equation}
we have
 \begin{eqnarray}
{\tilde S}_1&=&-{\tilde z_1}{\tilde z_3}+{\tilde z_4}(-{\tilde z_2}+{\tilde z_4})=\frac{-2y_1e^{y_4}+y_3+y_3e^{y_4}-2e^{y_4}y_3(y_4-y_2)}{2-2e^{y_4}} ,\nonumber\\
{\tilde S}_2&=&{\tilde z_2}-{\tilde z_4}+{\tilde z_3}({\tilde z_4}-{\tilde z_2})=-\frac{e^{2y_4}(1+e^{y_4}(-1+y_3))(-2y_1+y_3)}{2(e^{y_4}-1)^2} ,\nonumber\\
{\tilde S}_3&=&{\tilde z_1}=\frac{e^{-y_4}}{2}+y_2-y_4 ,\nonumber\\
{\tilde S}_4&=&{\tilde z_2}-{\tilde z_4}=\frac{e^{2y_4}(-2y_1+y_3)}{2e^{y_4}-2},
\end{eqnarray}
such that, they satisfy in the following Poisson brackets
\begin{eqnarray}
 \lbrace {\tilde S}_1,{\tilde S}_2\rbrace ={\tilde S}_4\;,\; \lbrace {\tilde S}_1,{\tilde S}_3\rbrace ={\tilde S}_3,\; \lbrace {\tilde S}_1,{\tilde S}_4\rbrace ={\tilde S}_4,\; \lbrace {\tilde S}_2,{\tilde S}_3\rbrace ={\tilde S}_4,
\end{eqnarray}
i.e., a Poisson bracket $ \lbrace {\tilde S}_i,{\tilde S}_j \rbrace ={ f}_{ij}^{\;\;k} {\tilde S}_k  $, where ${ f}_{ij}^{\;\;k}$ are the structure constants  of ${A_{4,9}^0.iv} $ and constants of motion  on the Lie group ${\bf A_{4,9}^0}$  as a phase space can be obtained as follows:
\begin{eqnarray}\label{bb14}
{\tilde I}_1=2({\tilde z_2}-{\tilde z_4})+d {\tilde z_1}=\frac{e^{2y_4}(2y_1-y_3)}{e^{y_4}-1}+d(\frac{e^{-y_4}}{2}+y_2-y_4),\nonumber\\
{\tilde I}_2=2({\tilde z_2}-{\tilde z_4})^2+(d {\tilde z_1})^2=\frac{e^{4y_4}(-2y_1+y_3)^2}{2(e^{y_4}-1)^2}+(d(\frac{e^{-y_4}}{2}+y_2-y_4))^2.
\end{eqnarray}
We see that in terms of the Darboux coordinates, these constants of motion are related to the physical systems.
Note that using  relations (\ref{2A511p}), (\ref{2A511})  and  (\ref{A4555}) one can find the following transformations between coordinates
$\{{\tilde z}_i\}$ and $\{z_i\}$:
\begin{equation}
{\tilde z}_1=-z_4,\;\;\;{\tilde z}_2=\frac{1}{z_1}-z_3,\;\;\;{\tilde z}_3=z_2,\;\;\;{\tilde z}_4=z_1.
\end{equation}
These are transformations between the Darboux coordinates on the Lie groups ${\bf A_{4,9}^0.iv}$ and ${\bf A_{4,9}^0}$  as phase spaces.
One can see that these transformations are preserved the canonical Poisson structure   (\ref{4b2})  and  (\ref{bb24}) but not transform the constants of
motions $\tilde{I}_1$ and $\tilde{I}_2$ (\ref{bb14}) to $I_1$ and $I_2$ of (\ref{4bb2}). In this sense these are not canonical transformations.\\
\smallskip

{\bf Example 2)} Lie bialgebra $(A_2\oplus A_2,(A_2\oplus A_2).vi)$ \cite{JAF2}:\\
 Consider the Lie group $\bf{A_2\oplus A_2} $ as a phase space and
 $\bf{(A_2\oplus A_2).vi}$ as symmetry Lie group of a Hamiltonian system.  For this example, the Darboux coordinates have
the following forms \cite{JAF}:
 \begin{eqnarray}\label{A5b2}
 z_1&=&\frac{e^{2x_3}x_4-e^{x_3} }{-1+e^{x_3}},\hspace*{2cm} z_2=x_2,\nonumber\\ z_3&=&e^{-x_3},\hspace*{2.5cm}\;\;\;\;\;\;\;\; z_4=\frac{x_2^2-x_1}{q x_2},
 \end{eqnarray}
  such that in this Darboux coordinates, the symplectic structure on the Lie group ${\bf A_2\oplus A_2} $  which has the following form \cite{JAF2}:
 \begin{eqnarray}
 \lbrace x_1,x_2\rbrace =q x_2,\;\;\lbrace x_1,x_4\rbrace =1+e^{-x3},
 \end{eqnarray}
 can be simplified as
{\begin{equation}
 \lbrace z_1,z_3\rbrace =1~,~~~~ \lbrace z_2,z_4\rbrace =1 .
\end{equation}
 In this way, we have the following forms for the dynamical functions $S^i$ according to \cite{JAF}
\begin{eqnarray}\label{A4}
S^1&=&q z_2 z_4=-x_1+x_2^2 ,\nonumber\\S^2&=&-a z_4=\frac{a x_1}{q x_2}-\frac{a x_2}{q},\nonumber\\S^3&=&-b z_3=-b e^{-x_3},\nonumber\\
S^4&=&-z_1 z_3=\frac{1-e^{x_3}x_4}{-1+e^{x_3}},
\end{eqnarray}
where $a, b$ and $q$ are nonzero arbitrary constants.  These functions   satisfy  in the following Poisson brackets
 \begin{eqnarray}
 \lbrace S^1,S^2\rbrace &=&q S^2\;,\; \lbrace S^3,S^4\rbrace =S^3و
\end{eqnarray}
i.e., a Poisson bracket $ \lbrace S^i,S^j \rbrace ={\tilde f}^{ij}_{\;\;k} S^k  $, where, ${\tilde f}^{ij}_{\;\;k}$ are  the
structure constants of  the symmetry Lie algebra  $(A_2\oplus A_2).vi$. The invariants of the above system
 are $(S^1, S^3)$,  $(S^2, S^4)$, $(S^1, S^4)$ or $(S^2, S^3)$ such that one can consider one of these $S^i$ as Hamiltonian of the integrable systems; in terms of
 Darboux coordinates these systems are physical once.
Now, by using (\ref{A3}) and Poisson structures on Lie group ${\bf A_2\oplus A_2}$ and ${\bf (A_2\oplus A_2).vi}$ one can find $x_i$'s as functions of $y_i$ as follows:
\begin{equation}\label{A5}
x_1=q(y_1- y_2)\;,\;\;x_2=-\frac{y_2}{q(y_2-y_1)}\;,\;\;x_3=y_4\;,\;\;x_4=-y_3,
\end{equation}
The Darboux coordinates on the Lie group ${\bf (A_2\oplus A_2).vi}$ have the following forms \cite{JAF}:
 \begin{eqnarray}\label{A5b1}
{\tilde z_1}&=&-\frac{e^{y_4}(1+e^{y_4} y_3)}{e^{y_4}-1},\hspace*{2cm} {\tilde z_2}=\frac{y_2 }{q (y_1-y_2)},\nonumber\\
 {\tilde z_3}&=&e^{-y_4},\hspace*{3cm}\;\;\;\;\;\;\;\;{\tilde z_4}= \frac{q^3(y_1-y_2)^3-y_2^2}{q^2 y_2(-y_1+y_2)},
 \end{eqnarray}
  such that in this Darboux coordinates, the symplectic structure on the Lie group ${\bf (A_2\oplus A_2).vi}$ which has the following form \cite{JAF2}:
 \begin{eqnarray}
 \lbrace y_1,y_2\rbrace =y_2,\;\;\lbrace y_3,y_4\rbrace =1-e^{y_4},\;\;
 \end{eqnarray}
 can be simplified as
{\begin{equation}
 \lbrace {\tilde z_1},{\tilde z_3}\rbrace =1~,~~~~ \lbrace {\tilde z_2},{\tilde z_4}\rbrace =1 .
\end{equation}
Now after using (\ref{A4})  and  substituting (\ref{A5}) and using  (\ref{5A2AA}) with
\begin{equation}
C=\left(
\begin{matrix}
q&0&0&0\cr
0&a&0&0\cr
0&0&0&b\cr
0&0&-1&0
 \end{matrix}
\right),
\end{equation}

we have
 \begin{eqnarray}
{\tilde S} _1&=&{\tilde z_2}{\tilde z_4}=\frac{-q^3(y_1-y_2)^3+y_2^2}{q^3(y_1-y_2)^2} ,\nonumber\\
{\tilde S} _2&=&-{\tilde z_4}=\frac{-q^3(y_1-y_2)^3+y_2^2}{q^2(-y_1+y_2)y_2},\nonumber\\{\tilde S} _3&=&{\tilde z_1}{\tilde z_3}=\frac{-1-e^{y_4}y_3}{-1+e^{y_4}} ,\nonumber\\
{\tilde S} _4&=&-{\tilde z_3}=-e^{-y_4}, \nonumber
\end{eqnarray}
such that, they satisfy the following Poisson brackets
 \begin{eqnarray}
 \lbrace {\tilde S}_1,{\tilde S}_2\rbrace &=&{\tilde S}_2\;,\; \lbrace {\tilde S}_3,{\tilde S}_4\rbrace ={\tilde S}_4,
\end{eqnarray}
i.e., a Poisson bracket $ \lbrace {\tilde S}_i,{\tilde S}_j \rbrace =f_{ij}^{\;\;k} {\tilde S}_k  $, where, $f_{ij}^{\;\;k}$ are  the structure constants of the symmetry Lie algebra  ${A_2\oplus A_2}$. The invariants of the above system
 are $({\tilde S}_1,{\tilde S}_3)$,  $({\tilde S}_2,{\tilde S}_4)$,  $({\tilde S}_1,{\tilde S}_4)$ or $({\tilde S}_2,{\tilde S}_3)$ such that one can consider one of these ${\tilde S}_i$ as Hamiltonian of the integrable systems, which in terms of the Darboux coordinates $\{ \tilde{z}_i\}$ its related to a physical systems.
 For this example using  (\ref{A5b1}), (\ref{A5})  and  (\ref{A5b2}) one can find the following trivial transformations between coordinates
${\tilde z}_i$ and $z_i$:
\begin{equation}
{\tilde z}_1=z_1,\;\;\;{\tilde z}_2=z_2,\;\;\;{\tilde z}_3=z_3,\;\;\;{\tilde z}_4=z_4.
\end{equation}
In this case these transformations preserved the canonical Poisson structures and transform  the constant of motions to each other
(($S^1\rightarrow q \tilde{S}_1; S^2\rightarrow a \tilde{S}_2; S^3\rightarrow b \tilde{S}_4; S^4\rightarrow -\tilde{S}_3$)
 $S^1$ transform to $\tilde{S}_1$ and $S^2$ transform to $\tilde{S}_2$ but  $S^3$ transform to $\tilde{S}_4$, and $S^4$ transform to $\tilde{S}_3)$; so these transformations are  canonical transformation. Note that these are
transformation between to phase space ${\bf (A_2\oplus A_2)}$ and ${\bf (A_2\oplus A_2).vi}$.
 \smallskip

{\bf Example 3)}  Lie bialgebra $(A_{4,9}^0,A_{4,9}^0.iv)$ \cite{JAF2}:\\
 Consider the Lie group $\bf{A_{4,9}^0} $ as a phase space and
 $\bf{A_{4,9}^0.iv}$ as symmetry Lie group of a Hamiltonian system.  For this example, the Darboux coordinates have
the following forms \cite{JAF}:
 \begin{eqnarray}\label{A5m32}
 z_1&=&-x_3,\hspace*{2cm} z_2=-\frac{e^{x_4}(-2e^{x_4}x_1+2e^{2x_4}x_1-x_3-e^{x_4}x_3+e^{2x_4}x_3-2e^{x_4}x_2x_3)}{2(-1+e^{x_4/2})},\nonumber\\ z_3&=&
 \frac{1+2e^{x_4}x_2}{2(-1+e^{x_4})},\hspace*{2cm}\;\;\;\;\;\;\;\; z_4=e^{-x_4},
 \end{eqnarray}
  such that in this Darboux coordinates, the symplectic structure on the Lie group $\bf{A_{4,9}^0} $  which has the following form \cite{JAF2}:
 \begin{eqnarray}
 \lbrace x_1,x_2\rbrace =\frac{1-e^{-2x_4}}{2},\;\;\lbrace x_1,x_3\rbrace =e^{-x_4}x_3,\;\;\lbrace x_1,x_4\rbrace =1-e^{-x_4},\;\;\lbrace x_2,x_3\rbrace =1-e^{-x_4},
 \end{eqnarray}
 can be simplified as
{\begin{equation}
 \lbrace z_1,z_3\rbrace =1~,~~~~ \lbrace z_2,z_4\rbrace =1 .
\end{equation}
 In this way, we have the following forms for the dynamical functions $S^i$ according to \cite{JAF}
\begin{eqnarray}\label{2A4}
S^1&=&z_1 z_3+z_2 z_4=\frac{2x_3+2e^{x_4}(x_1+2x_2x_3)-e^{2x_4}(2x_1+x_3+2x_2x_3)}{2(-1+e^{x_4})^2} ,\nonumber\\S^2&=&- z_3+z_2z_3\nonumber\\&=& -\frac{(1+2e^{x_4}x_2)(2+e^{x_4}(-4-x_3+e^{x_4}(2+2(-1+e^{x_4})x_1+(-1+e^{x_4}-2x_2)x_3)))}{4(-1+e^{x_4})^3} ,\nonumber\\S^3&=&- z_4=- e^{-x_4},\nonumber\\
S^4&=&-a z_3=\frac{a+2ae^{x_4}x_2}{2-2e^{x_4}},
\end{eqnarray}
where $a \in \mathbb{R} -\{0\}$. These functions  satisfy in the following Poisson brackets
\begin{eqnarray}
 \lbrace S^1,S^2\rbrace =S^4\;,\; \lbrace S^1,S^3\rbrace =S^3\;,\; \lbrace S^1,S^4\rbrace =S^4\;,\; \lbrace S^2,S^3\rbrace =S^4.
\end{eqnarray}
i.e., a Poisson bracket $ \lbrace S^i,S^j \rbrace ={\tilde f}^{ij}_{\;\;k} S^k  $, with ${\tilde f}^{ij}_{\;\;k}$ are  the
structure constants of  the symmetry Lie algebra  $A_{4,9}^0.iv$.
The invariants of the above system are $(S^3,S^4)$ or $(S^2,S^4)$, such that one can consider one of these $S^i$ as Hamiltonian of the integrable systems, for which in terms of the Darboux coordinates these are shown a physical system.
Now, by using (\ref{A3}) and Poisson structures on Lie group ${\bf A_{4,9}^0}$ and ${\bf A_{4,9}^0.iv}$ one can find $x_i$'s as functions of $y_i$ as follows:
\begin{equation}\label{2A5}
x_1=- y_4\;,\;\;x_2=y_1- y_3\;,\;\;x_3=y_2\;,\;\;x_4=y_1,
\end{equation}
For the Lie group ${\bf A_{4,9}^0.iv}$ as phase space the Darboux coordinates have the following forms \cite{JAF}:
 \begin{eqnarray}\label{A5m3}
{\tilde z_1}&=&-y_2,\hspace*{3cm} {\tilde z_2}=\frac{e^{y_1}(y_2(1-e^{2y_1}+e^{y_1}(1+2y_1-2y_3))+2e^{y_1}(e^{y_1}-1)y_4)}{2 (e^{y_1}-1)^2},\nonumber\\
 {\tilde z_3}&=&\frac{1+2e^{y_1}(y_1-y_3)}{2 (e^{y_1}-1)},\hspace*{3cm}\;\;\;\;\;\;\;\;{\tilde z_4}= e^{-y_1},
 \end{eqnarray}
  such that in this Darboux coordinates, the symplectic structure on the Lie group  ${\bf A_{4,9}^0.iv}$ which has the following form \cite{JAF2}:
 \begin{eqnarray}
 \lbrace y_1,y_4\rbrace =1-e^{-y_1},\;\;\lbrace y_2,y_3\rbrace =1-e^{-y_1},\;\;\lbrace y_2,y_4\rbrace =y_2e^{-y_1},\;\;\lbrace y_3,y_4\rbrace =\frac{1+e^{-2y_1}-2e^{-y_1}}{2},
 \end{eqnarray}
 can be simplified as
{\begin{equation}
 \lbrace {\tilde z_1},{\tilde z_3}\rbrace =1~,~~~~ \lbrace {\tilde z_2},{\tilde z_4}\rbrace =1 .
\end{equation}
Now after using  (\ref{2A4})  and  substituting (\ref{2A5}) and using  (\ref{5A2AA}) with
\begin{equation}
C=\left(
\begin{matrix}
0&0&0&-1\cr
1&0&-1&0\cr
0&1&0&0\cr
a&0&0&0
 \end{matrix}
\right),
\end{equation}

we have
 \begin{eqnarray}
{\tilde S} _1&=&-{\tilde z_3}=\frac{1+2e^{y_1}(y_1-y_3)}{2-2e^{y_1}} ,\nonumber\\
{\tilde S} _2&=&-{\tilde z_4}=-e^{-y_1},\nonumber\\
{\tilde S} _3&=&-{\tilde z_2}{\tilde z_3}=-\frac{e^{y_1}(1+2e^{y_1}(y_1-y_3))(-y_2+e^{2y_1}(y_2-2y_4)-e^{y_1}(y_2-2y_2(y_1-y_3)-2y_4)}{4(-1+e^{y_1})^3} ,\nonumber\\
{\tilde S} _4&=&-{\tilde z_1}{\tilde z_3}-{\tilde z_2}{\tilde z_4}=-\frac{2y_2+2e^{y_1}(2y_2(y_1-y_3)-y_4)-e^{2y_1}(y_2+2y_2(y_1-y_3)-2y_4)}{2(-1+e^{y_1})^2}, \nonumber
\end{eqnarray}
such that, they satisfy the following Poisson brackets
\begin{eqnarray}
 \lbrace {\tilde S}_1,{\tilde S}_4\rbrace ={\tilde S}_1\;,\; \lbrace {\tilde S}_2,{\tilde S}_3\rbrace ={\tilde S}_1
 ,\; \lbrace {\tilde S}_2,{\tilde S}_4\rbrace ={\tilde S}_2,
\end{eqnarray}
i.e., a Poisson bracket $ \lbrace {\tilde S}_i,{\tilde S}_j \rbrace ={ f}_{ij}^{\;\;k} {\tilde S}_k  $, with ${ f}_{ij}^{\;\;k}$ are  the
structure constants of  the symmetry Lie algebra  $A_{4,9}^0$.
The invariants of the above system
 are $({\tilde S}_1, {\tilde S}_3)$, $({\tilde S}_3, {\tilde S}_4)$ or $({\tilde S}_1, {\tilde S}_2)$ such that one can consider one of these ${\tilde S}_i$ a
 s Hamiltonian of the integrable systems, where these are  physical systems, in terms of  Darboux coordinates.
 For this example using  (\ref{A5m3}), (\ref{2A5})  and  (\ref{A5m32}) one can find the following transformations between coordinates
${\tilde z}_i$ and $z_i$:
\begin{equation}
{\tilde z}_1=z_1,\;\;\;{\tilde z}_2=z_2,\;\;\;{\tilde z}_3=z_3,\;\;\;{\tilde z}_4=z_4.
\end{equation}
In this case these transformations preserved the canonical Poisson structures and transform  the constant of motions to each other
(i.e., $S^1\rightarrow -{\tilde S}_4$; $S^2\rightarrow {\tilde S}_1-{\tilde S}_3$; $S^3\rightarrow {\tilde S}_2$; $S^4\rightarrow a{\tilde S}_1$); so these transformations are  canonical transformation. Note that these are
transformation between to phase space ${\bf A_{4,9}^0}$ and ${\bf A_{4,9}^0.iv}$.\\
 \smallskip

{\bf Example 4)} Lie bialgebra $(A_{4,9}^1,A_{4,9}^1.i)$ \cite{JAF2}:\\
 Consider the Lie group $\bf{A_{4,9}^1} $ as a phase space and
 $\bf{A_{4,9}^1.i}$ as symmetry Lie group of a Hamiltonian system.  For this example, the Darboux coordinates have
the following forms \cite{JAF}:
 \begin{eqnarray}\label{A5m42}
 z_1&=&-\frac{2e^{2x_4}x_2}{e^{2x_4}-1},\hspace*{1.5cm} z_2=
 \frac{e^{-4x_4}(-2 x_1+2e^{2x_4}x_1-2x_2x_3+e^{2x_4}x_2x_3)}{(-1+e^{2x_4})^2},\nonumber\\ z_3&=&
x_3 ,\hspace*{2cm}\;\;\;\;\;\;\;\; z_4=e^{-2x_4},
 \end{eqnarray}
  such that in this Darboux coordinates, the symplectic structure on the Lie group $\bf{A_{4,9}^1} $ which has the following form \cite{JAF2}:
 \begin{eqnarray}
 \lbrace x_1,x_2\rbrace =-\frac{x_2}{4},\;\;\lbrace x_1,x_3\rbrace =\frac{x_3-2e^{-2x_4}x_3}{4},
 \;\;\lbrace x_1,x_4\rbrace =-\frac{1}{4}(1-e^{-2x_4}),\;\;\lbrace x_2,x_3\rbrace =-\frac{1}{2}(1-e^{-2x_4}),
 \end{eqnarray}
 can be simplified as
{\begin{equation}
 \lbrace z_1,z_3\rbrace =1~,~~~~ \lbrace z_2,z_4\rbrace =1 .
\end{equation}
 In this way, we have the following forms for the dynamical functions $S^i$ according to \cite{JAF}
\begin{eqnarray}\label{3A4}
S^1&=&-1/4(2z_1 z_3+z_2 z_4)=-\frac{e^{2x_4}(2(e^{2x_4}-1)x_1+(2-3e^{2x_4})x_2x_3)}{4(-1+e^{x_4})^2} ,\nonumber\\S^2&=&- z_2z_3= -\frac{e^{2x_4}x_3(2(e^{2x_4}-1)x_1+(2-3e^{2x_4})x_2x_3)}{(-1+e^{x_4})^2}  ,\nonumber\\S^3&=&- z_4=- e^{-2x_4},\nonumber\\
S^4&=&- z_3=-x_3.
\end{eqnarray}
 such that these functions  satisfy in the following Poisson brackets
\begin{eqnarray}
 \lbrace S^1,S^2\rbrace =-\frac{1}{4}S^2\;,\; \lbrace S^1,S^3\rbrace =-\frac{1}{4}S^3\;,\; \lbrace S^1,S^4\rbrace =-\frac{1}{4}S^4\;,\; \lbrace S^2, S^3\rbrace =-S^4.
\end{eqnarray}
i.e., a Poisson bracket $ \lbrace S^i,S^j \rbrace ={\tilde f}^{ij}_{\;\;k} S^k  $, with ${\tilde f}^{ij}_{\;\;k}$ are  the
structure constants of  the symmetry Lie algebra  $A_{4,9}^1.i$.
The invariants of the above system
 are $(S^2,S^4)$ or $(S^3,S^4)$, such that one can consider one of these $S^i$ as Hamiltonian of the integrable systems; in terms of
  Darboux coordinates these are physical systems.
Now, by using (\ref{A3}) and Poisson structures on Lie group ${\bf A_{4,9}^1}$ and ${\bf A_{4,9}^1.i}$ one can find $x_i$'s as  functions of $y_i$ as follows:
\begin{equation}\label{3A5}
x_1=\frac{1}{4} y_4\;,\;\;x_2=- y_3\;,\;\;x_3=\frac{1}{4} y_2\;,\;\;x_4=-\frac{1}{4}y_1,
\end{equation}
For the Lie group ${\bf A_{4,9}^1.i}$ as phase space the Darboux coordinates have the following forms \cite{JAF}:
 \begin{eqnarray}\label{A5m4}
{\tilde z_1}&=&\frac{2 y_3}{1-e^{-\frac{y_1}{2}}},\hspace*{3cm} {\tilde z_2}=\frac{(2-e^{-\frac{y_1}{2}})y_2 y_3+2(e^{-\frac{y_1}{2}}-1)y_4}{4(e^{-\frac{y_1}{2}}-1)^2},\nonumber\\
 {\tilde z_3}&=&\frac{y_2}{4},\hspace*{3.3cm}\;\;\;\;\;\;\;\;{\tilde z_4}= e^{\frac{y_1}{2}},
 \end{eqnarray}
  such that in this Darboux coordinates, the symplectic structure on  $\bf{A_{4,9}^1.i} $ which has the following form \cite{JAF2}:
 \begin{eqnarray}
 \lbrace y_1,y_4\rbrace =4(-1+e^{\frac{y_1}{2}}),\;\;\lbrace y_2,y_3\rbrace =2(-1+e^{\frac{y_1}{2}}),\;\;
 \lbrace y_2,y_4\rbrace =y_2(-1+2e^{\frac{y_1}{2}}),\;\;\lbrace y_3,y_4\rbrace =y_3,
 \end{eqnarray}
 can be simplified as
{\begin{equation}
 \lbrace {\tilde z_1},{\tilde z_3}\rbrace =1~,~~~~ \lbrace {\tilde z_2},{\tilde z_4}\rbrace =1 .
\end{equation}
Now after using  (\ref{3A4})  and  substituting (\ref{3A5})
and using  (\ref{5A2AA}) with
\begin{equation}
C=\left(
\begin{matrix}
0&0&0&1/4\cr
0&0&1&0\cr
0&1&0&0\cr
1&0&0&0
 \end{matrix}
\right),
\end{equation}

 we have
 \begin{eqnarray}
{\tilde S} _1&=&-{\tilde z_3}=-y_2/4 ,\nonumber\\
{\tilde S} _2&=&-{\tilde z_4}=-e^{-y_1/2},\nonumber\\
{\tilde S} _3&=&-{\tilde z_2}{\tilde z_3}=-\frac{e^{-y_1/2}y_2((2e^{y_1/2}-1)y_2 y_3-2(-1+e^{y_1/2})y_4)}{16(-1+e^{y_1/2})^2},\nonumber\\
{\tilde S} _4&=&-2 {\tilde z_1}{\tilde z_3}-{\tilde z_2}{\tilde z_4}=-\frac{(2(e^{y_1/2}-1)y_4+(3-2e^{y_1/2})y_2y_3)}{4(-1+e^{y_1/2})^2}, \nonumber
\end{eqnarray}
such that, they satisfy the following Poisson brackets
 \begin{eqnarray}
 \lbrace {\tilde S}_1,{\tilde S}_4\rbrace =2{\tilde S}_1\;,\; \lbrace {\tilde S}_2,{\tilde S}_3\rbrace ={\tilde S}_1
 ,\; \lbrace {\tilde S}_2,{\tilde S}_4\rbrace ={\tilde S}_2\;,\; \lbrace {\tilde S}_3,{\tilde S}_4\rbrace ={\tilde S}_3,
\end{eqnarray}
i.e., a Poisson bracket $ \lbrace {\tilde S}_i,{\tilde S}_j \rbrace ={ f}_{ij}^{\;\;k} {\tilde S}_k  $, with ${ f}_{ij}^{\;\;k}$ are  the
structure constants of  the symmetry Lie algebra  $A_{4,9}^1$.
The invariants of the above system
 are $({\tilde S}_1,{\tilde S}_2)$ or $({\tilde S}_1,{\tilde S}_3)$, such that one can consider one of these ${\tilde S}_i$ as Hamiltonian of the integrable systems, which are  physical systems in terms of  Darboux coordinates.
 For this example using  (\ref{A5m4}), (\ref{3A5})  and  (\ref{A5m42}) one can find the following transformations between coordinates
${\tilde z}_i$ and $z_i$:
\begin{equation}
{\tilde z}_1=z_1,\;\;\;{\tilde z}_2=z_2,\;\;\;{\tilde z}_3=z_3,\;\;\;{\tilde z}_4=z_4.
\end{equation}
In this case these transformations preserved the canonical Poisson structures and transform the constant of motions to each other
(i.e., $S^1\rightarrow \frac{1}{4}{\tilde S}_4$; $S^2\rightarrow {\tilde S}_3$; $S^3\rightarrow {\tilde S}_2$; $S^4\rightarrow {\tilde S}_1$); so these transformations are  canonical transformation. Note that these are
transformation between to phase space ${\bf A_{4,9}^1}$ and ${\bf A_{4,9}^1.i}$.\\
 \smallskip

{\bf Example 5)} Lie bialgebra $(A_{4,7}.i,A_{4,7})$ \cite{JAF2}:\\
 Consider the Lie group $\bf{A_{4,7}.i} $ as a phase space and
 $\bf{A_{4,7}}$ as symmetry Lie group of a Hamiltonian system.  For this example, the Darboux coordinates have
the following forms \cite{JAF}:
 \begin{eqnarray}\label{A5m52}
 z_1&=&e^{-x_1},\hspace*{6cm} z_2=x_3,\nonumber\\ z_3&=&\frac{e^{2x_1}(2x_4-e^{x_1}(x_2^2+2x_2 x_3+2x_4))}{4(e^{x_1}-1)^2} ,\hspace*{.6cm}\;\;\;\;\;\;\;\; z_4=\frac{e^{x_1}x_2}{1-e^{x_1}},
 \end{eqnarray}
  such that in this Darboux coordinates, the symplectic structure on  $\bf{A_{4,7}.i} $ which has the following form \cite{JAF2}:
 \begin{eqnarray}
 \lbrace x_1,x_4\rbrace =2-2e^{-x_1},\;\;\lbrace x_2,x_3\rbrace =1-e^{-x_1},
 \;\;\lbrace x_2,x_4\rbrace =-x_2(1-2e^{-x_1}),\;\;\lbrace x_3,x_4\rbrace =x_2+x_3,
 \end{eqnarray}
 can be simplified as
{\begin{equation}
 \lbrace z_1,z_3\rbrace =1~,~~~~ \lbrace z_2,z_4\rbrace =1 .
\end{equation}
 In this way, we have the following forms for the dynamical functions $S^i$ according to \cite{JAF}
\begin{eqnarray}\label{4A4}
S^1&=&-z_3=\frac{e^{2x_1}(-2x_4+e^{x_1}(x_2^2+2x_2x_3+2x_4)}{4(-1+e^{x_1})^2} ,\nonumber\\
S^2&=&-z_2z_3=\frac{e^{2x_1}x_3(-2x_4+e^{x_1}(x_2^2+2x_2x_3+2x_4)}{4(-1+e^{x_1})^2} ,\nonumber\\
S^3&=& z_4=\frac{x_2 e^{x_1}}{1-e^{x_1}},\nonumber\\
S^4&=&(-2z_1+z_2^2/2)z_3-z_2z_4, \nonumber\\
&=& -\frac{e^{x_1}(8(x_2 x_3+x_4)+e^{2x_1 }x_3^2(x_2^2+2x_2x_3+2x_4)-2e^{x_1}(2x_2(x_2+4x_3)+(4+x_3^2)x_4))}{8(e^{x_1}-1)^2}.
\end{eqnarray}
   such that these functions satisfy in  the following Poisson brackets
\begin{eqnarray}
 \lbrace S^1,S^4\rbrace =2S^1\;,\; \lbrace S^2,S^3\rbrace =S^1\;,\; \lbrace S^2,S^4\rbrace =S^2\;,\; \lbrace S^3,S^4\rbrace =S^2+S^3.
\end{eqnarray}
i.e., a Poisson bracket $ \lbrace S^i,S^j \rbrace ={\tilde f}^{ij}_{\;\;k} S^k  $, with ${\tilde f}^{ij}_{\;\;k}$ are  the
structure constants of  the symmetry Lie algebra  $A_{4,7}$.
The invariants of the above system
 are $(S^1,S^2)$ or $(S^1,S^3)$, such that one can consider one of these $S^i$ as Hamiltonian of the integrable systems, where in terms of
  Darboux coordinates these are physical systems.
Now, by using (\ref{A3}) and Poisson structures on Lie group ${\bf A_{4,7}}$ and ${\bf A_{4,7}.i}$ one can  find $x_i$'s
as  functions of $y_i$ as   follows:
\begin{equation}\label{4A5}
x_1=2 y_4\;,\;\;x_2=y_2\;,\;\;x_3= y_3\;,\;\;x_4=\frac{4y_1-e^{2y_4}(4y_1+y_2^2+4y_2 y_3-y_3^2)+4y_2y_3}{2(e^{2y_4}-1)},
\end{equation}
For the Lie group ${\bf A_{4,7}}$ as phase space the Darboux coordinates have the following forms \cite{JAF}:
 \begin{eqnarray}\label{A5m5}
{\tilde z_1}&=& e^{-2y_4},\hspace*{8cm} {\tilde z_2}=y_3,\nonumber\\
 {\tilde z_3}&=&\frac{e^{4 y_4}(4(-1+e^{2y_4}) y_1-y_3(-2(-2+e^{2y_4}) y_2+e^{2y_4}y_3))}{4(e^{2y_4}-1)^2},
 \hspace*{0.5cm}\;{\tilde z_4}=\frac{e^{2y_4}y_2}{1-e^{2y_4}},
 \end{eqnarray}
  such that in this Darboux coordinates, the symplectic structure on  $\bf{A_{4,7}} $ which has the following form \cite{JAF2}:
 \begin{eqnarray}
 \lbrace y_1,y_2\rbrace =\frac{y_2-y_3}{2},\;\;\lbrace y_1,y_3\rbrace =y_3(e^{-2y_4}-\frac{1}{2}),\;\;
 \lbrace y_1,y_4\rbrace =\frac{1}{2}(1-e^{-2y_4}),\;\;\lbrace y_2,y_3\rbrace =1-e^{-2y_4},
 \end{eqnarray}
 can be simplified as
{\begin{equation}
 \lbrace {\tilde z_1},{\tilde z_3}\rbrace =1~,~~~~ \lbrace {\tilde z_2},{\tilde z_4}\rbrace =1 .
\end{equation}
Now after using  (\ref{4A4})  and  substituting (\ref{4A5})
and using  (\ref{5A2AA}) with
\begin{equation}
C=\left(
\begin{matrix}
0&0&0&2\cr
0&0&-1&0\cr
0&1&0&0\cr
-2&0&0&0
 \end{matrix}
\right),
\end{equation}
we have:
 \begin{eqnarray}
{\tilde S} _1&=&{\tilde z_1}{\tilde z_3}-1/4 {\tilde z_2}({\tilde z_2}{\tilde z_3}-2{\tilde z_4})\nonumber\\
&=&-\frac{e^{2y_4}(4(e^{2y_4}-1)y_1(-4+e^{2y_4}y_3^2)+y_3(4e^{2y_4}y_3-e^{4y_4}y_3^2+2y_2(4-2e^{2y_4}y_3^2+e^{4y_4}y_3^2)))}{16(e^{2y_4}-1)^2} ,\nonumber\\
{\tilde S} _2&=&{\tilde z_4}=\frac{e^{2y_4}y_2}{1-e^{2y_4}},\nonumber\\
{\tilde S} _3&=&{\tilde z_2}{\tilde z_3}=-\frac{e^{4y_4}y_3(-4(-1+e^{2y_4})y_1+y_3(-2(-2+e^{2y_4})y_2+e^{2y_4}y_3))}{4(e^{2y_4}-1)^2},\nonumber\\
{\tilde S} _4&=&-{\tilde z_3}/2=-\frac{e^{4y_4}(-4(-1+e^{2y_4})y_1+y_3(-2(-2+e^{2y_4})y_2+e^{2y_4}y_3))}{8(e^{2y_4}-1)^2}, \nonumber
\end{eqnarray}
such that, they satisfy the following Poisson brackets
 \begin{eqnarray}
 \lbrace {\tilde S}_1,{\tilde S}_2\rbrace =\frac{1}{2}{\tilde S}_2-\frac{1}{2}{\tilde S}_3\;,\; \lbrace {\tilde S}_1,{\tilde S}_3\rbrace =\frac{1}{2}{\tilde S}_3
 ,\; \lbrace {\tilde S}_1,{\tilde S}_4\rbrace ={\tilde S}_4\;,\; \lbrace {\tilde S}_2,{\tilde S}_3\rbrace =2{\tilde S}_4\;,
\end{eqnarray}
i.e., a Poisson bracket $ \lbrace {\tilde S}_i,{\tilde S}_j \rbrace ={ f}_{ij}^{\;\;k} {\tilde S}_k  $, with ${ f}_{ij}^{\;\;k}$ are  the
structure constants of  the symmetry Lie algebra  $A_{4,7}.i$.
The invariants of the above system
 are $({\tilde S}_2,{\tilde S}_4)$ or $({\tilde S}_3,{\tilde S}_4)$, such that one can consider one of these ${\tilde S}_i$
  as Hamiltonian of the integrable systems,  which are  physical systems in terms of  Darboux coordinates.
 For this example using  (\ref{A5m5}), (\ref{4A5})  and  (\ref{A5m52}) one can find the following transformations between coordinates
${\tilde z}_i$ and $z_i$:
\begin{equation}
{\tilde z}_1=z_1,\;\;\;{\tilde z}_2=z_2,\;\;\;{\tilde z}_3=z_3,\;\;\;{\tilde z}_4=z_4.
\end{equation}
In this case these transformations preserved the canonical Poisson structures and transform the constant of motions to each other
(i.e., $S^1\rightarrow 2{\tilde S}_4$; $S^2\rightarrow -{\tilde S}_3$; $S^3\rightarrow {\tilde S}_2$; $S^4\rightarrow -2{\tilde S}_1$); so these transformations are  canonical transformation. Note that these are
transformation between to phase space ${\bf A_{4,7}}$ and ${\bf A_{4,7}.i}$.\\
 \smallskip


\end{document}